\documentclass[11pt,twocolumn,twoside,a4paper,amsmath,amssymb,jnpcs,apsmy,showkeys,showpacs]{revtex4-2}

\usepackage[cp1251]{inputenc}
\usepackage[russian]{babel}
\usepackage{jnpcs}
\usepackage{graphics} 
\usepackage{epsfig}
\usepackage{amsfonts,amssymb,latexsym}

\DeclareGraphicsExtensions{.jpg,.pdf,.mps,.png} 
\firstpage{21} \nlpage{\pageref{last}} \nvolume{25} \nnumber{2}
\nyear{2022}
\def\nfpage{\thepage}
\renewcommand{\theequation}{\thesection.\theequation}
\numberwithin{equation}{section}

\newcommand{\pd}[2]{\frac{\partial #1}{\partial #2}}

\begin{document}

\chead[S.~O. Komarov, A.~K. Gorbatsievich,  A.~S. Garkun, and G.~V.
Vereshchagin]{Electromagnetic radiation and electromagnetic self-force of a point charge in the vicinity of Schwarzschild black hole}

\title{Electromagnetic radiation and electromagnetic self-force of a point charge in the vicinity of Schwarzschild black hole}

\author{S. O. Komarov}
\email{StasKomarov@tut.by}
\affiliation{Belarusian State University, Department of Theoretical Physics and Astrophysics, Nezavisimosti Av. 220030 Minsk, Belarus\\
ICRANet-Minsk, National Academy of Sciences of Belarus, Nezavisimosti Av. 68-2, 220072 Minsk, Belarus}

\author{A. K. Gorbatsievich}
\email{Gorbatsievich@bsu.by}
\affiliation{Belarusian State University, Department of Theoretical Physics and Astrophysics, Nezavisimosti Av. 220030 Minsk, Belarus}

\author{A. S. Garkun}
\email{Garkun@iaph.bas-net.by}
\affiliation{Institute of Applied Physics, National Academy of Sciences of Belarus, Academic st. 16, 220072 Minsk, Belarus}

\author{G. V. Vereshchagin}
\email{Veresh@icra.it}
\affiliation{ICRANet, Piazza della Repubblica, 10, 65122 Pescara, Italy\\ 
ICRANet-Minsk, National Academy of Sciences of Belarus, Nezavisimosti Av. 68-2, 220072 Minsk, Belarus\\
INAF — IAPS, Via del Fosso del Cavaliere, 100, 00133 Rome, Italy}


\begin{abstract}
Point charge, radially moving in the vicinity of a black hole is considered. Electromagnetic field in wave zone and in the small neighbourhood of the charge is calculated. Numerical results of the calculation of the spectrum of electromagnetic radiation of the point charge are presented. Covariant approach for the calculation of electromagnetic self-force is used for the case of the slowly moving charge. Numerical results for the self-force in the case of slow motion of the particle are obtained and compared to the results in literature.
\end{abstract}

\received{7 November, 2022}

\pacs{04.25.dg, 04.70.Bw}

\keywords{Electrodynamics in qurves space-time, motion in external gravitational field}


\maketitle

\section{Introduction }

The problem of point charge motion in external gravitational field of a black hole is very important for the understanding of the mechanisms of plasma accretion e. g. within the recently proposed unified model of gamma-ray bursts and quasars \cite{Ruffini2019IE,Ruffini2019,Ruffini2020}. There are many open questions about the electromagnetic radiation and the motion of such charges  \cite{Shatski-NovikovEN,Tursunov2018,Poisson2011}. The main difficulties emerge in the description of  electromagnetic radiation reaction and in the description of the influence of external gravitational field on electromagnetic field distribution and on the motion of the charge. 

In this paper we consider electromagnetic field of point charge moving radially in the vicinity of a Schwarzschild black hole. By using the approach that is based on the papers \cite{Ruffini1971,Ruffini1972} we study analytically and numerically electromagnetic spectrum of radiation of the charge in wave zone.  Also we consider electromagnetic field and electromagnetic self-force for the charge. Using covariant formalism for the description of electromagnetic field from the paper \cite{Poisson2011} we obtain analytical formula for the self-force that can be used in vicinity of the event horizon of black hole. We obtain numerical results and compare it with those in the literature. The results are discussed in "Conclusions" section.    

Latin indices have values from 1 to 4, Greek indices from 1 to 3. The signature of the metric is +2. We use the system of units where the speed of light in vacuum is  $c=1$.

\section{Formulation of the problem}

Consider a point charge near spherically symmetric black hole. Metric of the spherically symmetric (Schwarzschild) black hole has the following form \cite{Stephani}:
\begin{eqnarray}
&&\mathrm{d}s^2=\frac{\mathrm{d}r^2}{1-2M/r}+r^2\mathrm{d}\theta^2+\notag\\
&&r^2\sin^2\theta\mathrm{d}\varphi^2-\left(1-\frac{2M}{r}\right)\mathrm{d}t^2\,.
\end{eqnarray} 
Here $x^i=\{r,\,\theta,\,\varphi,\,t\}$ are Schwarzschild coordinates. $M=Gm_{BH}$, where $G$ is the gravitational constant and $m_{BH}$ is the mass of the black hole.

Electromagnetic field $F_{ij}$ of the charged particle can be found from the Maxwell equations in curved space-time:
\begin{eqnarray}\label{Maxwell}
&&F^{ls}{}_{;s}=4\pi j^l\,;\label{1M}\\
&&F_{ij;k}+F_{ki;j}+F_{jk;i}=0\,.\label{2M}
\end{eqnarray}
where $j^l$ is four-current vector:
\begin{eqnarray}\label{current}
&&j^l(x^k)=eu^l(x^{\alpha'}(x^4),x^4)\times\\
&&\frac{\delta(x^1-x^{1'}(x^{4'}))\delta(x^2-x^{2'}(x^{4'}))\delta(x^3-x^{3'}(x^{4'}))}{\sqrt{-g}u^4(x^{\alpha'}(x^4),x^4)}\notag\,.
\end{eqnarray}

It follows from equation (\ref{2M}) that electromagnetic field tensor $F_{ij}$ can be expressed in the form 
\begin{equation}
F_{ls}=A_{l;s}-A_{s;l}\,,
\end{equation}
where $A_k$ 4-potential of electromagnetic field.

\section{Radial motion of the particle}
Electromagnetic energy radiated by a test charge, falling radially into the black hole, can be found following the papers \cite{Ruffini1971,Ruffini1972}. For this purpose we are looking for the solution of (\ref{Maxwell}) in the form of multipole expansion \cite{Ruffini1971}:
\begin{eqnarray}\label{anzats}
&&A_1=\sum\limits_{l=0,\,m=0}^{\infty}h_{lm}(r,t)Y_{lm}(\theta,\phi)\,;\notag\\
&&A_2=\sum\limits_{l=0,\,m=0}^{\infty}k_{lm}(r,t)\pd{Y_{lm}(\theta,\phi)}{\theta}\,;\notag\\
&&A_3=\sum\limits_{l=0,\,m=0}^{\infty}k_{lm}(r,t)\pd{Y_{lm}(\theta,\phi)}{\phi}\,;\notag\\
&&A_4=\sum\limits_{l=0,\,m=0}^{\infty}f_{lm}(r,t)Y_{lm}(\theta,\phi)\,.
\end{eqnarray}
Here, $Y_{lm}(\theta,\phi)$ are spherical functions.

\begin{figure}
	\centering
	\includegraphics[angle=0, width=\columnwidth]{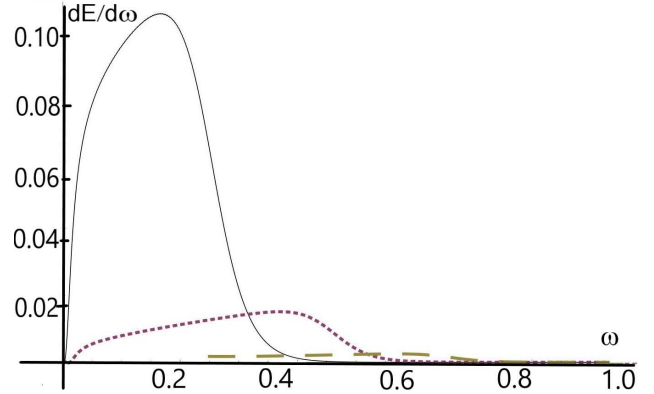}
	\caption{Spectrum of electromagnetic radiation of the particle, falling radially into the Schwarzschild black hole, for different multipole index $l$: for $l=1$ (solid), for $l=2$ (point), for $l=3$ (dotted). Frequency is calculated in units of $1/M$. } \label{dEdo2}
\end{figure}

Then, for the multipole components of $A_k$ obtain:
\begin{eqnarray}\label{maineq}
&&(g^{11}b_{lm}{}_{,r})_{,r}+g^{44}b_{lm,44}-\frac{l(l+1)}{r^2}b_{lm}=\notag\\
&&\frac{1}{l(l+1)}\left((r^2\psi_{lm})_{,r}-(r^2\eta_{lm})_{,4}\right)\,;\notag\\
&&b_{lm}=\frac{r^2}{l(l+1)}(h_{lm}{}_{,r}-f_{lm}{}_{,r})\,;\notag\\
&&k_{lm}{}_{,4}-f_{lm}=g^{11}b_{lm}{}_{,4}-\frac{r^2}{l(l+1)}\psi_{lm}\,;\notag\\
&&h_{lm}-k_{lm}{}_{,r}=g^{44}b_{lm}{}_{,4}+\frac{r^2}{l(l+1)}\eta_{lm}\,.
\end{eqnarray}
Here functions $\psi$ and $\eta$ are multipole coefficients for the components of electric current $j_k$:
\begin{eqnarray}
&&4\pi j_1=\sum\limits_{l,m=0}^{\infty}\eta_{lm}(r,t)Y_{lm}(\theta,\phi)\,;\notag\\
&&4\pi j_4=\sum\limits_{l,m=0}^{\infty}\psi_{lm}(r,t)Y_{lm}(\theta,\phi)\,.\notag
\end{eqnarray}
Introduce the Fourier components of the mentioned functions by using the following relations:
\begin{eqnarray}
&\psi_{lm}=\int\limits^{+\infty}_{-\infty}\tilde{\psi}_{lm}e^{-i\omega t}\mathrm{d}\omega\,;\notag\\
&\eta_{lm}=\int\limits^{+\infty}_{-\infty}\tilde{\eta}_{lm}e^{-i\omega t}\mathrm{d}\omega\,;\notag\\
&b_{lm}=\int\limits^{+\infty}_{-\infty}\tilde{b}_{lm}e^{-i\omega t}\mathrm{d}\omega\,;\notag
\end{eqnarray}
Then, equation for $b_{lm}$  has the following form:
\begin{eqnarray}
&\left[\left(1-\frac{2M}{r}\right)\tilde{b}_{lm,r}\right]_{,r}+\left(\frac{\omega^2}{1-2M/r}-\frac{l(l+1)}{r^2}\right)\tilde{b}_{lm}=\notag\\
&a(r)\,.\label{finalspectreq}
\end{eqnarray}
Here we introduce function $a(r)$:  
\begin{eqnarray}
&a(r)=\dfrac{q}{2\pi}\dfrac{2l+1}{l(l+1)}e^{it(r)}\times\notag\\
&\left[\dfrac{ME/r^2}{(E^2-1+2M/r)^{3/2}}-\dfrac{i\omega}{E^2-1+2M/r}\right]\,.\notag
\end{eqnarray}
Solution of equation (\ref{finalspectreq}) must satisfy boundary conditions for radiation. They are require the following asymptotics at the event horizon and at infinity \cite{MembraneParadigmEN}: 
\begin{eqnarray}
&b_{lm}(r,\,\omega)\rightarrow e^{i\omega r}\,,\text{ for }r\rightarrow +\infty\,,\notag\\
&b_{lm}(r,\,\omega)\rightarrow e^{-i\omega r*}\,,\text{ for }r\rightarrow 2M\,.\label{boundaryconditions}
\end{eqnarray}
Here $r*=r+2M\ln{(r-2M)}$ is the turtle coordinate. Using Green's function technique, we obtain solution of (\ref{finalspectreq}) in the form:
\begin{eqnarray}
&b_{lm}(r,\,\omega)=\dfrac{1}{W(r)(1-2M/r)}\times\notag\\
&\left[y_1(r)\int\limits^{+\infty}_{r}y_2(r')a(r')\mathrm{d}r'+\right.\notag\\
&\left.y_2(r)\int\limits^{r}_{2M}y_1(r')a(r')\mathrm{d}r'\right]\,.
\end{eqnarray}
Here $y_1(r)$ and $y_2(r)$ --- linear independent solutions of the homogeneous equation with respect to (\ref{finalspectreq}), $W(r)$ is Wronsky determinant:
\begin{equation*}
W(r)=y_1(r)y'_2(r)-y'_1(r)y_2(r)\,.
\end{equation*}
In order to satisfy boundary conditions (\ref{boundaryconditions}) it is necessary to choose solutions $y_1$ as ingoing wave at the event horizon, and $y_2$ as outgoing wave at the spatial infinity. First solution can be written by using confluent Heun functions for variable $\xi=\frac{r}{2M}$ (see, e. g. \cite{FizievHeunF,MaierHeunF}):
\begin{equation}\label{form}
y_1=\xi^{2}(\xi-1)^{-2i\omega}e^{-2i\omega\xi}H(\xi)\,.
\end{equation} 
In Wolfram Mathematica it has the following representation:
\begin{equation*}
H(\xi)=HeunC[q,\,\alpha,\,\gamma,\,\delta,\,\varepsilon,\xi]\,,
\end{equation*}
where
\begin{eqnarray}
&q=l(l+1)-2\,;\notag\\
&\alpha=-8i\omega\,;\notag\\
&\gamma=3\,;\notag\\
&\varepsilon=-4i\omega\,;\notag\\
&\delta=1-4i\omega\,.\notag
\end{eqnarray}
Then, energy radiated by the point charge can be calculated from Fourier coefficients $b(\omega,r)$:
\begin{equation}
\frac{\mathrm{d}E}{\mathrm{d}\omega}=\frac{i}{4}\sum\limits_{l=1}^{\infty}l(l+1)\tilde{b}_{,r}(\omega,r)\tilde{b}^{*}(\omega,r)\,.
\end{equation}
Results of numerical calculation of the spectrum of radiation are presented on Fig. \ref{dEdo2}.
\section{General solution for electromagnetic field and self-force}
\begin{figure}
	\leavevmode
	\centering
	\includegraphics[angle=0, width=\columnwidth]{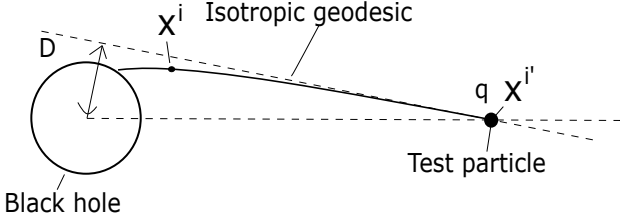}
	\caption{Isotropic geodesic connecting the particle in $x^{i'}$ and the point of observation $x^i$.} \label{ChargeEn}
\end{figure}

General solution for electromagnetic potential $A^k$ can be found using the formalism of Synge's world function (see, e. g. \cite{Poisson2011}):
\begin{equation}\label{Apotential}
A^l=\left.q\frac{u^{l'}g^{l}{}_{l'}\sqrt{\Delta}}{\sigma_{k'}u^{k'}}\right|_{\sigma=0}+e\int\limits^{\tau'}_{-\infty}V^l{}_{l''}u^{l''}\mathrm{d}\tau''\,.
\end{equation}
Here $q$ is the charge of the particle, $\sigma$ is the Synge's world function:
\begin{eqnarray}
&\sigma(x^k,\,x^{k'})=\frac{1}{2}(\lambda_1-\lambda_0)\int\limits_{\lambda_0}^{\lambda_1}g_{ij}t^it^j\mathrm{d}\lambda=\notag\\
&\frac{1}{2}(\lambda_1-\lambda_0)^2||t^j||^2\,,\notag
\end{eqnarray}
$\lambda\in[\lambda_0,\,\lambda_1]$ is affine parameter on isotropic geodesic between $x^k$ and $x^{k'}$ (see fig. \ref{ChargeEn}).  
\begin{equation}\label{Apotential1}
A^l=\left.e\frac{u^{l'}g^{l}{}_{l'}\sqrt{\Delta}}{\sigma_{k'}u^{k'}}\right|_{\sigma=0}+e\int\limits^{\tau'}_{-\infty}V^l{}_{l''}u^{l''}\mathrm{d}\tau''\,.
\end{equation}
Here $V_{ii'}$ is a bitensor satisfying d'Alamber's equation in curved space-time:
\begin{equation}\label{eqV1}
g^{ij}\frac{D^2}{Dx^iDx^j}V^l{}_{l''}-R^l{}_sV^s{}_{l''}=0\,,
\end{equation}
and initial condition
\begin{equation}\label{eqV2}
V_{l'l'}=\frac{1}{12}R g_{l'l'}-\frac{1}{2}R_{l'l'}\,;
\end{equation}	
We use the "Method of images":
\begin{equation}\label{selfforce}
\left.q\frac{u^{l'}g^{l}{}_{l'}\sqrt{\Delta}}{\sigma_{k'}u^{k'}}\right|_{\sigma=0, r=2M}=\frac{q}{2r'}\,.
\end{equation} 
This formalism gives us possibilities to calculate electromagnetic self-force $F^r$ for the charged particle, fixed far from Schwarzschild black hole:
\begin{equation}\label{self-force}
F^r\approx q^2 \int\limits_{-\infty}^{0}V_{ji';r}u^{i'}g^j{}_{k'}u^{k'}\mathrm{d}\tau'\approx \frac{q^2 M}{r'^3}\,.
\end{equation}
This result coincides with those obtained in literature, where different approaches have been used (see e. g. \cite{McGruder1978,Vilenkin}).

\begin{figure}
	\leavevmode
	\centering
	\includegraphics[angle=0, width=\columnwidth]{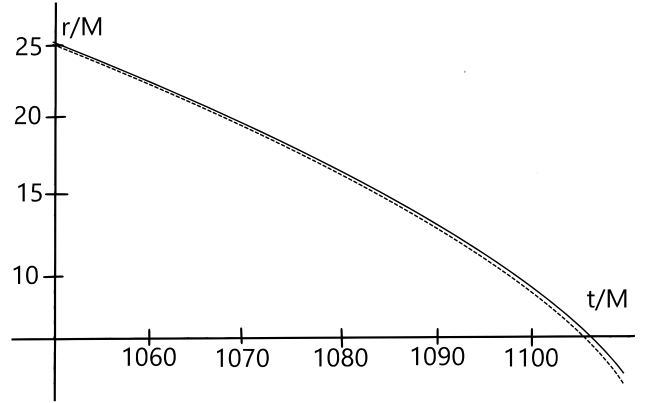}
	\caption{Radial coordinate of a test particle falling into the Schwarzschild black hole as function of its proper time with self-force (solid), and without it (dotted). The particle start at rest at time $t=0$. Here $q^2=0.1 M m$, where $m$ is the mass of the particle.} \label{ChargeMotion}
\end{figure}

It follows from the approach for the calculation of self-force that the obtained formula depends only on the electric field distribution in the space and due to this must be valid not only for fixed point charge but for the moving charge as well.  Numerical results  for the test particle under influence of the self-force (\ref{self-force}) and without self-force are presented in Fig. \ref{ChargeMotion}.

\section{Conclusions}
In the present work we show that the solution for the spectrum of the electromagnetic radiation of the charge falling into Schwarzschild black hole can be presented in compact form by using the confluent Heun functions. The numerical results of calculation show that most of the energy radiated is determined by the first ($l=1$) multipole moment in expansion. This is in agreement with the results of previous papers, where only numerical solution of homogeneous equation for $b_{lm}$ in limited region of space was calculated (see, e. g. \cite{Ruffini1971,Ruffini1972}). 

We found analytical and numerical solution for electromagnetic self-force of a charge moving in the vicinity of Schwarzschild black hole using covariant approach that is based on the results of the paper \cite{Poisson2011}.  Unlike post-Newtonian consideration in other papers \cite{McGruder1978,Vilenkin}, the presented approach gives us possibilities to obtain analytical expression for the electromagnetic self-force that is valid in the small vicinity of the event horizon of Schwarzschild black hole. For the point charge far from black hole numerical and analytical results are in agreement with post-Newtonian consideration (see, e. g. \cite{McGruder1978,Vilenkin}).  

\section*{Acknowledgment}
The work was supported by BRFFR Foundation in the framework of the F21ICR BRFFR-ICRANet project.

\label{last}
\end{document}